\documentclass[aps,
  pra,                
  reprint,
  superscriptaddress      
]{revtex4-2}

\usepackage{amsmath}
\usepackage{graphicx}
\usepackage{bm}
\usepackage{mathrsfs}
\usepackage{lineno}
\usepackage{setspace}
\usepackage{comment}
\usepackage{scalerel}
\usepackage{amssymb}  

\begin{document}

\title{A Reduced Action Integral for Photon--Photon Interactions in Vacuum}

\author{D. Ramsey}
\email{dram@lle.rochester.edu}
\affiliation{
University of Rochester, Laboratory for Laser Energetics, Rochester, New York 14623-1299, USA
}
\author{M. S. Formanek}
\affiliation{ELI Beamlines Facility, The Extreme Light Infrastructure ERIC, 252 41 Doln\'{i} B\v{r}e\v{z}any, Czech Republic}
\author{J. P. Palastro} 
\affiliation{
University of Rochester, Laboratory for Laser Energetics, Rochester, New York 14623-1299, USA
}

\date{\today}

\newcommand{\unitvec}[1]{\ensuremath{\bm{\hat{\mathrm{#1}}}}}
\newcommand{\bvec}[1]{\ensuremath{\bm{\mathrm{#1}}}}
\newcommand{\conjugate}{\mkern-12mu\raisebox{0.3ex}{$^*$}}
\newcommand{\smallparallel}{{\scalebox{0.7}{$\scriptstyle\parallel$}}}

\begin{abstract}
Electromagnetic waves propagating through vacuum can polarize virtual electron–positron pairs; this polarization, in turn, nonlinearly modifies their propagation. A semi-classical nonlinear wave equation describing the propagation is derived from the Euler--Heisenberg Lagrangian density, which captures vacuum polarization effects up to the one-loop level. Here, we present a reduced-action-integral approach that enables rapid modeling of nonlinear phenomena arising from the Euler--Heisenberg Lagrangian. Application of the variational principle to the reduced action provides equations of motion for familiar light-pulse parameters, such as spot size, phase, polarization, and phase-front curvature, without requiring full-field simulations. Three examples demonstrate the utility of the approach: phase modulation, birefringence, and frequency mixing.
\end{abstract}
\maketitle
\newpage

\section{Introduction}
Homogeneous solutions to Maxwell's equations obey the superposition principle. For classical fields, this means that electromagnetic waves do not interact in vacuum. Quantum electrodynamics, in contrast, predicts that photons interact in vacuum by polarizing virtual electron--positron pairs. An effect of this prediction, namely photon--photon scattering, has been observed in interactions mediated by strong Coulombic fields \cite{atlas2017evidence, sirunyan2019evidence, aad2019observation, brandenburg2023report}. Meanwhile, experiments involving interactions between purely electromagnetic waves have placed an upper bound on the cross section for photon--photon scattering \cite{bernard2000search}, but have yet to produce an observable signature. This has motivated proposals for geometries \cite{FFvacBire, Martin} and experiments \cite{di2006light, ReviewKing, king2016vacuum, MP3Workshop, gonoskov2022charged, fedotov2023advances, ahmadiniaz2025towards, Hans&Antonino} aimed at detecting the effects of photon--photon scattering in the interaction of electromagnetic waves. The proposed experiments include collisions between bright x-ray beams and high-intensity optical pulses \cite{ahmadiniaz2025towards}, as well as the interaction of multiple high-intensity optical pulses \cite{Hans&Antonino}---the latter being the flagship experiment of the planned NSF OPAL laser \cite{nsf_opal_flagships}.

At photon energies much lower than the electron rest mass energy, photon--photon scattering can be described using a semi-classical approach. The leading-order nonlinear corrections to classical electrodynamics are described by the Euler--Heisenberg Lagrangian \cite{EulerHeisen,Schwinger},
\begin{equation}\label{eq:EHL}
\begin{split}
        &\mathcal{L}_\mathrm{EH} = \frac{1}{8\pi}\left( |\bvec{\mathcal{E}}|^2-|\bvec{\mathcal{B}}|^2\right)\\
        +&\frac{\alpha}{360 \pi^2E_\mathrm{S}^2}\left[2\left( |\bvec{\mathcal{E}}|^2-|\bvec{\mathcal{B}}|^2\right)^2+7\left(\bvec{\mathcal{E}}\cdot\bvec{\mathcal{B}}\right)^2\right],
\end{split}
\end{equation}
where the first term is the classical electromagnetic Lagrangian density \cite{Jackson} and the second term  encodes the one-loop quantum correction due to virtual electron--positron pairs. Here, $\bvec{\mathcal{E}}$ and $\bvec{\mathcal{B}}$ are the electric and magnetic fields, respectively, $\alpha \approx 1/137$ is the fine-structure constant, $E_\mathrm{S}  = \alpha e/r_e^2 = 4.41\times10^{13} \,\mathrm{statV/cm}
$ ($1.32\times10^{18} \,\mathrm{V/m}$ in SI units) is the Schwinger field, $r_e$ is the classical electron radius, and $e$ the elementary charge. The correction term gives rise to a vacuum polarization density $\bvec{\mathcal{P}} = \partial\mathcal{L}_\mathrm{EH}/\partial{\bvec{\mathcal{E}}} - \bvec{\mathcal{E}}/4\pi$ and magnetization density $\bvec{\mathcal{M}} = \partial\mathcal{L}_\mathrm{EH}/\partial{\bvec{\mathcal{B}}} +\bvec{\mathcal{B}}/4\pi$, revealing a third-order nonlinear response:
\begin{equation}\label{eq:Dens}
\begin{split}
    \bvec{\mathcal{P}} &= \frac{\xi}{4\pi}[2(|\bvec{\mathcal{E}}|^2-|\bvec{\mathcal{B}}|^2)\bvec{\mathcal{E}} + 7(\bvec{\mathcal{E}}\cdot\bvec{\mathcal{B}})\bvec{\mathcal{B}}],\\
    \bvec{\mathcal{M}} &= \frac{\xi}{4\pi}[2(|\bvec{\mathcal{B}}|^2-|\bvec{\mathcal{E}}|^2)\bvec{\mathcal{B}} + 7(\bvec{\mathcal{B}}\cdot\bvec{\mathcal{E}})\bvec{\mathcal{E}}],\\
\end{split}
\end{equation}
where $\xi = \alpha/(45\pi E_\mathrm{S}^2)$. 

\begin{figure*}[t]
\centering
\includegraphics[width=1\textwidth]{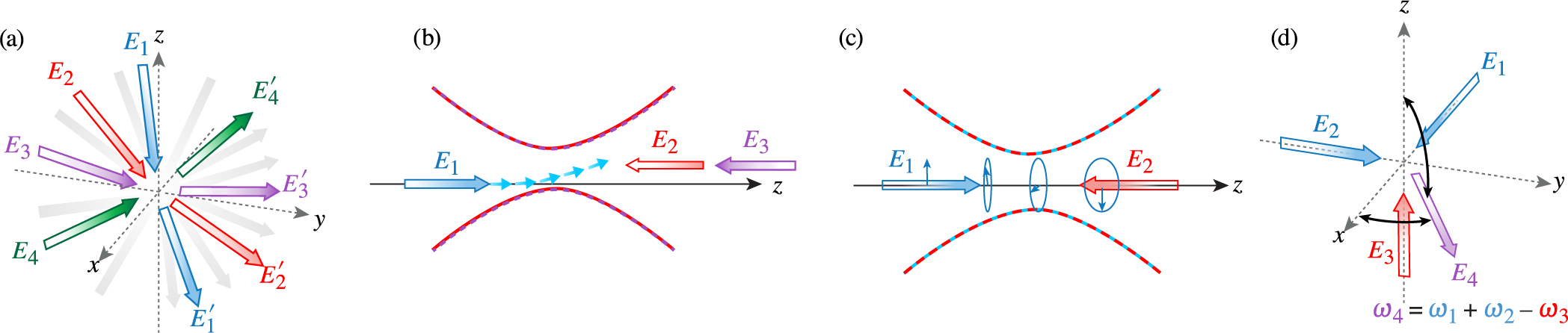}
\caption{(a) The third-order nonlinear response $(\bvec{\mathcal{P}}, \bvec{\mathcal{M}})$ of a polarized virtual-pair medium couples multiple input pulses (unprimed). Several processes can modify the outbound pulses (primed): (b) Phase modulation induced by one or more pulses ($E_2$, $E_3$) can deflect another pulse ($E_1$). The blue arrow chain depicts the resulting centroid trajectory of $E_1$ as it deviates from its initial heading. (c) The relative polarization of $E_1$ and $E_2$ affects the strength of their coupling, resulting in a birefringence that modifies the polarization of $E_1$. For instance, an initially linearly polarized field $E_1$ can become elliptically polarized, as illustrated by the blue ellipses. (d) Three spectrally distinct pulses ($E_1$, $E_2$, and $E_3$) mix to generate a fourth pulse ($E_4$) at a new frequency ($\omega_4$).}\label{fig:1}
\end{figure*}

As in a classical medium with a third-order nonlinearity, electromagnetic waves traversing the quantum vacuum (i.e., a ``medium'' of virtual electron--positron pairs) can interact through several nonlinear processes. This is illustrated in Fig. \ref{fig:1} for (a) the most general configuration, where multiple input waves interact to produce multiple output waves, as well as the specific processes of (b) phase modulation, (c) birefringence, and (d) frequency mixing \cite{klein1964birefringence, di2006light, lundstrom2006using, Lundin, heinzl2006observation, VarApproach1D, ReviewAntonino, ReviewKing, king2016vacuum, PhysRevA.98.023817, jeong2020photon,  grismayer2021quantum, ahmadiniaz2025towards, Martin, wang2024exploring, Limitations, ZhangAddsSemiClassical, matheron2025simulating}. The vector nature of the polarization and magnetization densities gives rise to many distinct interaction geometries for each process. Thus, maximizing signatures of these processes requires exploration of a vast parameter space. Given the extremely small magnitude of the nonlinearity--- $\alpha|\bvec{\mathcal{E}}|^2/(180\pi^2 E_\mathrm{S}^2) \sim \mathcal{O}(10^{-12})$ for even the most powerful laser systems \cite{OneStrongLaser,bromage2019technology, radier202210,nsf_opal_capabilities}---thorough exploration of potential geometries and parameters is crucial for identifying detectable signatures and guiding experimental design.

Here, we employ a reduced-action approach to model the evolution of familiar light-pulse parameters for any number of electromagnetic pulses interacting in vacuum. The approach enables rapid modeling of the nonlinear phenomena arising from the Euler--Heisenberg Lagrangian, which can be used to identify promising configurations for experiments. In the approach, nonlinear effects manifest as changes in the centroid, wavevector, spot size, phase, polarization, or power of each pulse. The electromagnetic field of each pulse is represented in an action integral using physically motivated trial functions parameterized by these quantities. Equations of motion for each parameter are obtained by integrating the action over the transverse coordinates and applying the principle of least action. This approach, also called Rayleigh–Ritz optimization, is a standard method for modeling nonlinear interactions \cite{anderson1979variational,anderson1983variational,duda2000variational,shankar2012principles, landau2013quantum}, which has been previously applied to the Euler--Heisenberg Lagrangian for one-dimensional interactions \cite{VarApproach1D}. Here, we consider three-dimensional interactions. The result is a streamlined framework expressed in terms of intuitive, experimentally observable quantities that can pinpoint geometries and parameters with detectable signatures and discard those without.

\begin{figure*}[t]
\centering
\includegraphics[width=1\textwidth]{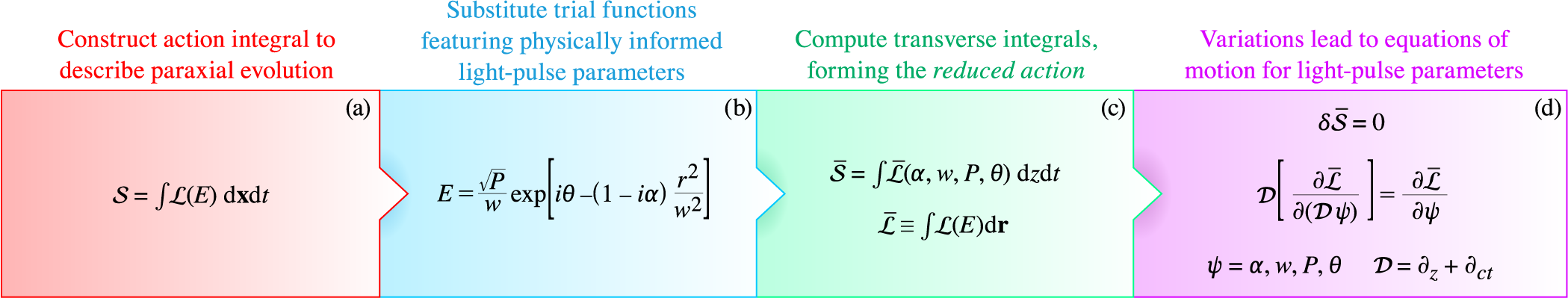}
\caption{The reduced-action-integral approach: (a) Construct an action integral for the nonlinear, paraxial evolution of all spectrally distinct or spatiotemporally disjoint fields. (b) Substitute physically motivated trial functions for the fields, parameterized by familiar light-pulse quantities, such as the spot size, power (amplitude), and phase. (c) Integrate over the coordinates orthogonal to the propagation direction to obtain a reduced action integral. (d) Apply the principle of least action to derive equations for the evolution of the pulse quantities.} \label{fig:2}
\end{figure*}

The panels of Fig. \ref{fig:2} outline the approach and structure of the article:
\begin{enumerate}
    \item The approach begins with the inhomogeneous wave equation describing the evolution of the electromagnetic field in vacuum. The slowly varying envelope and paraxial approximations are applied to construct a Lagrangian density $\mathcal{L}$ and action integral $\mathcal{S}$ for a set of spectrally distinct or spatiotemporally disjoint field components $E$, representing a system with infinite degrees of freedom for each coordinate in space and time [Fig. \ref{fig:2}(a), Sec. \ref{sec:SVEA}]. 
    \item In the action integral, each field component is approximated using a trial function featuring physically motivated light-pulse parameters [Fig. \ref{fig:2}(b), Sec. \ref{sec:ReAct}].
    \item Integrating the Lagragian density over the coordinates transverse to the propagation direction of each field component reduces the infinite transverse degrees of freedom to a finite set of light-pulse parameters, forming a reduced action integral $\bar{\mathcal{S}}$ [Fig. \ref{fig:2}(c), Sec. \ref{sec:ReAct}].
    \item Applying the principle of least action yields Euler--Lagrange equations, which can be recast as equations of motion for the light-pulse parameters. The formalism is then used to model the three nonlinear processes illustrated in Fig. 1 [Fig. \ref{fig:2}(d), Sec. \ref{sec:Apps}].
\end{enumerate}
Finally, Sec. \ref{sec:Con} summarizes the results and describes an extension to arbitrary interaction geometries.

\section{The paraxial action integral}\label{sec:SVEA}
In any medium with polarization density $\bvec{\mathcal{P}}$ and magnetization density $\bvec{\mathcal{M}}$, the propagation of  electromagnetic waves is described by the inhomogeneous wave equations:
\begin{equation}\label{eq:EwaveEQ}
\begin{split}
    &\left(\nabla^2-\frac{1}{c^2}\frac{\partial^2}{\partial t^2}\right)\bvec{\mathcal{E}} \\
    &=4 \pi \left[\frac{1}{c^2}\frac{\partial^2 \bvec{\mathcal{P}}}{\partial t^2}  + \frac{1}{c}\nabla \times \frac{\partial \bvec{\mathcal{M}}}{\partial t}  -\nabla(\nabla \cdot \bvec{\mathcal{P}})\right],
\end{split}
\end{equation}
\begin{equation}\label{eq:BwaveEQ}
\begin{split}
    &\left(\nabla^2-\frac{1}{c^2}\frac{\partial^2}{\partial t^2}\right)\bvec{\mathcal{B}}  \\
    &= 4 \pi \left[\nabla^2 \bvec{\mathcal{M}}-\frac{1}{c}\nabla\times \frac{\partial \bvec{\mathcal{P}}}{\partial t}-\nabla(\nabla\cdot\bvec{\mathcal{M}})\right],
\end{split}
\end{equation}
where $c$ is the speed of light in vacuum.
Assume the total electromagnetic field $(\bvec{\mathcal{E}},\bvec{\mathcal{B}})$, as well as the polarization and magnetization densities $( \bvec{\mathcal{P}},\bvec{\mathcal{M}})$, each consist of $N$ spectrally distinct or spatiotemporally disjoint components, as illustrated in Fig. \ref{fig:1}(a). Each component can be expressed as an envelope multiplied by a rapidly oscillating carrier:
\begin{equation}
    \bvec{\mathcal{E}}= \sum_{\beta=1}^N \frac{1}{2}\bvec{E}_\beta(\bvec{x},t) \,\mathrm{exp}[i (\bvec{k}_\beta\cdot \bvec{x} - \omega_\beta t)] +\mathrm{c.c.},
\end{equation}
with analogous expressions for the magnetic field, polarization density, and magnetization density. The rapid phase of component $\beta$ is $ \bvec{k}_\beta\cdot \bvec{x} - \omega_\beta t$, where $\omega_\beta$ is the frequency and $\bvec{k}_\beta$ is the wavevector, with magnitude $k_\beta = \omega_\beta/c$ in the unit direction $\unitvec{k}_\beta$. The set of $N$ spectrally distinct or spatiotemporally disjoint field components excludes components that do not satisfy the vacuum dispersion relation, $\omega^2 = c^2 k_\beta^2$, as such modes cannot propagate to the far field.

If the envelope of each component varies over spatiotemporal scales much longer than its wavelength and period, the evolution of the full electromagnetic field can be approximated by that of the envelopes. This is equivalent to making the slowly varying and paraxial approximation for each component. Equations~\eqref{eq:EwaveEQ} and \eqref{eq:BwaveEQ} then reduce to $2N$ coupled equations---one for each envelope and its complex conjugate:
\begin{equation}\label{eq:INHOMVec}
\begin{split}
   & \left[\nabla^2_{\perp \beta} +2ik_\beta\left(\hat{\bvec{k}}_\beta\cdot\nabla+\frac{\partial}{\partial (ct)}\right)\right]{\bvec{E}}_\beta   \\
   & = - 4 \pi k_\beta^2 \left( {\bvec{P}}_\beta  - \unitvec{k}_\beta \times {\bvec{M}}_\beta \right),
\end{split}
\end{equation}
where $\nabla_{\perp \beta}^2$ is the transverse Laplacian in the plane perpendicular to the propagation direction $\unitvec{k}_\beta$, and the magnetic field envelopes can be found from ${\bvec{B}}_\beta = \unitvec{k}_\beta\times{\bvec{E}}_\beta$. For the inhomogeneity, only the leading-order terms proportional to $k_\beta^2$ are retained, while the paraxial approximation $\bvec{k}_\beta \cdot \bvec{E}_\beta \approx 0$ justifies neglecting the term $\nabla(\nabla \cdot \bvec{P}_\beta) \approx 0$. 

To model general interactions among components, particularly in geometries that give rise to birefringence, it is convenient to decompose the vector envelopes into their scalar projections. The resulting scalar envelopes are described using three labels:
\begin{enumerate}
    \item  Greek letters label spectrally distinct or spatiotemporally disjoint components (e.g., $\beta,\gamma,\mu,\nu$) ranging from 1 to $N$.
    \item  Latin letters label the envelope projection (e.g., $j,m,n,o$) onto one of the three Cartesian directions $(\unitvec{x},\, \unitvec{y},\,\unitvec{z})$. Note that while the propagation of each component is approximated as paraxial ($\mathbf{k}_\beta \cdot \bvec{E}_\beta \approx 0$), three projections may still be required when $\mathbf{k}_\beta$ is not aligned with the longitudinal unit vector of the global coordinate system $\unitvec{z}$.
    \item The envelope conjugation status, indicating whether an envelope or its complex conjugate is involved, is represented using Gothic letters (e.g., $\mathfrak{a},\mathfrak{b},\mathfrak{c},\mathfrak{d}$) and will appear as a superscript. The conjugation label will be $*$ when denoting the conjugate and is omitted otherwise. When used algebraically, it will be $-1$ for conjugated envelopes and $+1$ otherwise.
    \end{enumerate}

Equation \eqref{eq:Dens} shows that the nonlinear response described by the Euler--Heisenberg Lagrangian is third order with respect to the fields. Thus, envelopes coupled through the inhomogeneity form a ``phase-matched quartet.'' A phase-matched quartet consists of four scalar  envelopes, $(E_{\beta j}^\mathfrak{a},\,E_{\gamma m}^\mathfrak{b},\,E_{\nu n}^\mathfrak{c},\,E_{\mu o}^\mathfrak{d})$, with wavevectors and frequencies that satisfy the momentum and energy conservation relations
\begin{equation}\label{eq:defQuart}
\begin{split}
    \mathfrak{a}\bvec{k}_\beta +\mathfrak{b} \bvec{k}_\gamma +\mathfrak{c} \bvec{k}_\mu+\mathfrak{d} \bvec{k}_\nu &= 0,\\
    \mathfrak{a} \omega_\beta +\mathfrak{b} \omega_\gamma +\mathfrak{c} \omega_\mu +\mathfrak{d} \omega_\nu &= 0.
\end{split}
\end{equation}
The set of all distinct, order-independent, phase-matched quartets in a given system is denoted by $\mathcal{Q}$. 

Using the scalar-envelope notation, the inhomogeneity ${\bvec{P}}_\beta  - \unitvec{k}_\beta \times {\bvec{M}}_\beta$ arising from the Euler--Heisenberg Lagrangian can be expressed in terms of a third-order susceptibility tensor $\mathrm{X}^{(3)}$ acting on the product of three envelopes: 
\begin{equation}\label{eq:SourceTerm}
\begin{split}
&(\bvec{P}_\beta- \unitvec{k}_\beta \times \bvec{M}_\beta)_j\\
&=\sum_{\mathcal{Q}(E_{\beta j}^*)} \mathrm{X}^{(3)}_{jmno}(\beta,\gamma,\mu,\nu) E_{\gamma m}^\mathfrak{b} E_{\mu n}^\mathfrak{c} E_{\nu o}^\mathfrak{d},
\end{split}
\end{equation}
where $\mathrm{X}^{(3)}$ is defined in Appendix A, and the summation runs over all distinct phase-matched quartets that include $E_{\beta j}^*$, denoted by $\mathcal{Q}(E_{\beta j}^*)$. A notable property of $\mathrm{X}^{(3)}$ is that it vanishes when the wavevectors are parallel $\unitvec{k}_\beta=\unitvec{k}_\gamma=\unitvec{k}_\mu=\unitvec{k}_\nu\Rightarrow \mathrm{X}^{(3)} = 0$. This implies that co-propagating waves cannot polarize vacuum, and thus do not interact. In addition, within the paraxial approximation, a single field cannot interact with itself, which precludes self-phase modulation, self-focusing, and third-harmonic generation.

The dynamical system of envelopes can be encoded in an action integral: 
\begin{equation}\label{eq:FullactINT}
\mathcal{S} = \int \mathcal{L}(\bvec{E}_1,\partial\bvec{E}_1,\bvec{E}_2,\partial\bvec{E}_2,\ldots)\,\mathrm{d}^3\bvec{x}\,\mathrm{d}t,
\end{equation}
where $\mathcal{L}$ is the Lagrangian density and $\partial \bvec{E}$ denotes all space and time derivatives of an envelope. The stationary-action principle, $\delta \mathcal{S}= 0$, is applied over the infinite degrees of freedom inherent to each space--time coordinate $(\bvec{x},\,t)$, yielding Eq. \eqref{eq:INHOMVec}. The Lagrangian density separates into two terms,
\begin{equation}
    \mathcal{L} =\mathcal{L}_{\mathrm{PX}}+\mathcal{L}_{\mathrm{NL}}.
\end{equation}
The first term, $\mathcal{L}_\mathrm{PX}$, describes the paraxial propagation of each scalar envelope
\begin{equation}\label{eq:PXLag}
\begin{split}
    \mathcal{L}_{\mathrm{PX}} \equiv \sum_{\beta =1}^N \sum_{j = 1}^3 &\frac{1}{k_\beta^2}|\nabla_{\perp \beta}E_{\beta j}|^2 \\
   & + \frac{i}{k_\beta} (E_{\beta j}\mathcal{D}_\beta E_{\beta j}^* -E_{\beta j}^* \mathcal{D}_\beta E_{\beta j}),
    \end{split}
\end{equation}
where $\mathcal{D}_\beta \equiv \unitvec{k}_\beta\cdot\nabla+\partial_{ct}$ is the advection operator. The second term, $\mathcal{L}_{\mathrm{NL}}$, describes the nonlinear interaction among envelopes
\begin{equation}\label{eq:NLLag}
    \mathcal{L}_{\mathrm{NL}} =-4\pi \sum_{\mathcal{Q}} 
     \mathrm{X}^{(3)}_{jmno}(\beta,\gamma,\mu,\nu) E_{\beta j}^\mathfrak{a} E_{\gamma m}^\mathfrak{b} E_{\mu n}^\mathfrak{c} E_{\nu o}^\mathfrak{d},
\end{equation}
where the summation runs over the entirety of $\mathcal{Q}$.

Functional variations of $\mathcal{L}$ with respect to any envelope yield the corresponding inhomogeneous paraxial wave equation for its conjugate. For example, varying with respect to $E_{\beta j}^*$ yields the projection of Eq. \eqref{eq:INHOMVec} in direction $j$:
\begin{equation}\label{eq:PaXWav}
\begin{split}
&\left(\nabla^2_{\perp \beta} +  2ik_\beta\mathcal{D}_\beta\right)E_{\beta j} = \frac{\partial \mathcal{L}_{\mathrm{NL}}}{\partial E_{\beta j}^*}.
\end{split}
\end{equation}
Although the paraxial action model described by Eqs.~\eqref{eq:FullactINT}--\eqref{eq:PaXWav} enables direct analogy with well-known nonlinear optical processes (Fig.~\ref{fig:1}), it does not reduce the spatiotemporal degrees of freedom and thus offers little simplification in determining parameters or geometries suitable for their experimental detection.

\section{The reduced action integral}\label{sec:ReAct}
To form an action integral with reduced degrees of freedom, trial functions for each envelope are substituted into the action integral. The trial functions are expressed in terms of familiar light-pulse parameters, such as spot size, amplitude, and phase-front curvature, that can vary in time and position along the propagation axis. The transverse dependence is kept explicit, allowing the Lagrangian density to be integrated over coordinates orthogonal to the propagation axis. This projects the infinite transverse degrees of freedom onto a finite set, yielding a \emph{reduced action integral}. The variational derivatives are taken with respect to the light-pulse parameters, providing their equations of motion. 

The trial functions are chosen to resemble solutions of the homogeneous paraxial wave equation. Their selection should reflect the symmetries of the interaction, or absence thereof, as well as the relevant physical processes. For instance, Laguerre--Gaussian (LG) modes may be a natural choice for interactions that are or are nearly cylindrically symmetric, while Hermite--Gaussian (HG) modes may be preferable for geometries with lateral asymmetries. Using these modes as a basis, an envelope $E_{\beta j}$ can be expressed by either of the following superpositions:
\begin{widetext}
\begin{equation}\label{eq:LGmodes}
    E_{\beta j}(\bvec{x},t) = \sum_{p,\ell} 
    A_{p\ell}(z,t)\left(\frac{\sqrt{2}r}{w(z,t)}\right)^{|\ell|}L_p^{|\ell|}\left( \frac{2r^2}{w^2(z,t)}\right)\, \mathrm{exp}\left[-(1-i\alpha(z,t))\frac{r^2}{w^2(z,t)}+i\ell \phi+i\theta_{p\ell}(z,t)\right],
\end{equation}
or
\begin{equation}\label{eq:HGmodes}
\begin{split}
    E_{\beta j}(\bvec{x},t) =\sum_{p,q} 
    A_{pq}(z,t)\,&\mathrm{exp}\big[i\theta_{pq}(z,t) +i\bvec{k}_\perp(z,t)\cdot\big(\bvec{r}-\bvec{r}_\mathrm{c}(z,t)\big)\big]\\ 
    &\times H_{p}\left[\frac{\sqrt{2} \big(x-x_\mathrm{c}(z,t)\big)}{w_{x}(z,t)}\right]\mathrm{exp}\left[-(1-i\alpha_{x}(z,t))\frac{\big(x-x_\mathrm{c}(z,t)\big)^2}{w^2_{x}(z,t)}\right]\\
    &\times H_{q}\left[\frac{\sqrt{2} \big(y-y_\mathrm{c}(z,t)\big)}{w_{y}(z,t)}\right]\mathrm{exp}\left[-(1-i\alpha_{y}(z,t))\frac{\big(y-y_\mathrm{c}(z,t)\big)^2}{w^2_{y}(z,t)}\right],
\end{split}
\end{equation}
\end{widetext}
where $z$ is the coordinate along the propagation direction $\unitvec{k}$, $\bvec{r}=(x,y)$ is a vector perpendicular to $\unitvec{k}$ with $|\bvec{r}|= r$, and $\mathrm{tan}(\phi) = y/x$. The function $L_p^{|\ell|}$ is the generalized Laguerre polynomial with radial index $p$ and azimuthal index $\ell$, while $H_q$ is the Hermite polynomial of order $q$. The light-pulse parameters $w$,  $\alpha$,  $A$, and $\theta$ are real quantities that characterize the transverse pulse width (spot size), phase front curvature, amplitude, and phase, respectively. To account for lateral asymmetries, the HG modes are expressed in terms of independent light-pulse parameters in the $x$- and $y$-directions, denoted by subscripts $x$ and $y$ (e.g., $w_x$, $w_y$). The HG modes also feature light-pulse parameters that describe an evolving centroid, $\bvec{r}_\mathrm{c} = (x_\mathrm{c},\, y_\mathrm{c})$, and transverse wavevector, $\bvec{k}_\perp = (k_x,\, k_y)$. For notational simplicity, the indexing of the light-pulse parameters as well as $\bvec{r}$ and $z$ by the envelope labels $(\beta j)$ is suppressed, although each envelope $E_{\beta j}$ carries its own independent set of parameters and coordinates. From here on, the dependence of the light-pulse parameters on $(z,t)$ will be assumed but not explicitly written. 

The next step is to substitute the trial functions into $\mathcal{L}_{\mathrm{PX}}$ and $\mathcal{L}_{\mathrm{NL}}$ [Eqs. \eqref{eq:PXLag} and $\eqref{eq:NLLag}$], and then integrate over the transverse coordinates of each envelope, $\bvec{r}_\beta$. Evaluating the integrals of $\mathcal{L}_{\mathrm{PX}}$ yields
\begin{equation}\label{eq:PXReducedGen}
    \int{\mathcal{L}}_\mathrm{PX} \;\mathrm{d}\bvec{x}\mathrm{d}t =\sum_{\beta =1}^N \sum_{j = 1}^3\int\overline{\mathcal{L}}_{\mathrm{PX}}^{\beta j}\;\mathrm{d}z_{\beta} \mathrm{d}t, 
\end{equation}
where the overlined Lagrangian density $\overline{\mathcal{L}}_{\mathrm{PX}}^{\beta j}$ (now a line density) depends on the light-pulse parameters of envelope $E_{\beta j}$, as well as their longitudinal and temporal derivatives. To analytically evaluate the integrals of $\mathcal{L}_\mathrm{NL}$ over $\bvec{r}_{ \beta}$ for all interactions involving a pulse $E_{\beta j}$, the other pulses must either be nearly counter-propagating or have trial functions with light-pulse parameters that do not change during the interaction. Otherwise, the integrals must be computed perturbatively (see Sec. \ref{sec:Con} for discussion). For the analytically tractable cases of interest [Figs. \ref{fig:1}(b)--\ref{fig:1}(d)], a global coordinate system, $(x,y,z)$, can be defined allowing the transverse integral of $\mathcal{L}_{\mathrm{NL}}$ to be evaluated:
\begin{equation}\label{eq:NLReducedGen}
    \int{\mathcal{L}}_\mathrm{NL} \;\mathrm{d}^3\bvec{x}\mathrm{d}t = \int  \overline{\mathcal{L}}_{\mathrm{NL}}\;\mathrm{d}z \mathrm{d}t,
\end{equation}
where the overline indicates that these terms depend on the light-pulse parameters of the interacting pulses. Summing Eqs. \eqref{eq:PXReducedGen} and \eqref{eq:NLReducedGen} provides the total reduced action integral,
\begin{equation}\label{eq:reducedAct}
    \overline{\mathcal{S}} \equiv   \int  \left[\Bigg(\sum_{\beta =1}^N \sum_{j = 1}^3\overline{\mathcal{L}}_{\mathrm{PX}}^{\beta j} \Bigg) + \overline{\mathcal{L}}_{\mathrm{NL}}\right]\mathrm{d}z \mathrm{d}t, 
\end{equation}
where the integrand will be referred to as the reduced Lagrangian, $\overline{\mathcal{L}}$. 

Dynamic equations for the light-pulse parameters are obtained by requiring $\overline{\mathcal{S}}$ to be stationary, i.e., $\delta \overline{\mathcal{S}} = 0$. This condition is enforced by taking functional variations of $\overline{\mathcal{S}}$ with respect to the light-pulse parameters, leading to an Euler–-Lagrange equation for each parameter:
\begin{equation}\label{eq:ExEOMLG}
\mathcal{D}_\beta \left[\frac{ \partial \overline{\mathcal{L}}_\mathrm{PX}^{\beta j} }{\partial({\mathcal{D}_\beta} \psi_{\beta j g h})}\right]=\frac{\partial }{\partial \psi_{\beta j g h}} \left(\overline{\mathcal{L}}_\mathrm{PX}^{\beta j} + \overline{\mathcal{L}}_\mathrm{NL}\right),
\end{equation}
where $(g,h)$ represent the modal indices, either $(p,\ell)$ for LG modes or $(p,q)$ for HG modes, and $\psi_{\beta j gh}$ is any light-pulse parameter, e.g., $\psi_{\beta j p \ell} = A_{\beta j p \ell},\, \theta_{\beta j p \ell},\, w_{\beta j},\, \mathrm{or} \; \alpha_{\beta j}$ for the LG modes. The result of the Euler--Lagrange equations can then be recast into equations of motion for the light-pulse parameters.

The subsequent analysis focuses on interactions among pulses that remain well described by the lowest-order transverse modes. Thus, the transverse modal indices on $A$ and $\theta$ will be dropped. The study of processes that mix or involve higher-order modes is deferred to future work. In addition, due to the coupling between light-pulse parameters, it is convenient to express $A_{\beta j}$ in terms of the power,
\begin{equation}
P_{\beta j} \equiv \frac{c}{16}\times
\begin{cases} 
\ A_{\beta j}^2 w_{\beta j}^2,  \text{ LG},\\[2mm]
A_{\beta j }^2 w_{x,\beta j} w_{y,\beta j}, \text{ HG}
\end{cases}.
\end{equation}
When expressed in terms of $P_{\beta j}$, the reduced Lagrangian is rescaled by the constant factor $c/(16\pi)$ to remove extraneous coefficients. As a result, the reduced Lagrangian will have units of power, i.e., energy per unit time.

\section{Examples}\label{sec:Apps}
Three examples are presented below to illustrate the versatility of the reduced action integral in modeling nonlinear processes arising from the Euler--Heisenberg Lagrangian: phase modulation, birefringence, and frequency mixing. Within the reduced-action framework, these processes alter the light-pulse parameters. The modifications to propagation described by these parameters could serve as experimental observables to verify the predictions of the Euler--Heisenberg Lagrangian. 
A key advantage of the framework is that it yields relatively simple equations that help identify parameters and geometries with experimentally measurable effects while ruling out those that do not.

\subsection{Phase Modulation}
The Euler--Heisenberg Lagrangian predicts that one pulse can modify the phase and trajectory of another through a process known as cross-phase modulation \cite{VarApproach1D, di2006light, heinzl2006observation, ReviewAntonino, ReviewKing, wang2024exploring}. In the paraxial action integral, this interaction arises from quartets of the form $(E_{\beta j},\,E_{\beta j}^*,\,E_{\gamma m},\,E_{\gamma m}^*)$, which contribute a Kerr-like nonlinearity $\mathcal{L}_\mathrm{NL}\propto-|E_{\beta j}|^2|E_{\gamma m}|^2$. As a specific example, consider the three pulses illustrated in Fig.~\ref{fig:1}(b): a forward-propagating pulse ($\beta = 1$) and two backward-propagating, non-overlapping pulses ($\gamma = 2$ and $\mu = 3$), with $\unitvec{k}_1 = -\unitvec{k}_2= -\unitvec{k}_3= \unitvec{z}$. Each pulse acquires a spatially dependent phase due to the nonuniform intensity profiles of the other pulses, which alters the evolution of its spot size and centroid.

To isolate these effects from those of birefringence, the backward-propagating pulses are assigned polarization vectors that are parallel to each other and orthogonal to the forward-propagating pulse. The relevant couplings are $(E_{1 1},\,E_{1 1}^*,\,E_{22},\,E_{22}^*)$ and $(E_{1 1},\,E_{1 1}^*,\,E_{32},\,E_{32}^*)$. For brevity, the polarization labels will be omitted: ($E_{1 1}$, $E_{22}$, $E_{32}$) $\rightarrow$ ($E_1$, $E_{2}$, $E_3$). Under these conditions, the interaction  Lagrangian is
\begin{equation}
{\mathcal{L}}_\mathrm{NL}=-14\xi \left(|E_1|^2|E_2|^2+|E_1|^2|E_3|^2\right).
\end{equation}
Recall that because $\unitvec{k}_2 = \unitvec{k}_3=-\unitvec{z}$, the backward-propagating pulses do not couple.

To capture the centroid dynamics resulting from the interaction, the lowest-order HG mode is used as the trial function for each pulse
\begin{equation}\label{eq:simpTrial}
\begin{split}
    &E = A\,\mathrm{exp}\big[i\theta +i\bvec{k}_\perp\cdot\big(\bvec{r}-\bvec{r}_\mathrm{c}\big)\big]\\
                  & \times \exp{\left[-(1-i \alpha_x)\frac{(x-x_\mathrm{c})^2}{w_x^2}-(1-i \alpha_y)\frac{(y-y_\mathrm{c})^2}{w_y^2}\right]}.
\end{split}
\end{equation}
Substituting these trial functions into Eq. \eqref{eq:FullactINT} and following the procedure outlined in the previous section yields the reduced paraxial Lagrangian: 
\begin{equation}\label{eq:simpTrialLHG}
\begin{split}
    \overline{\mathcal{L}}_\mathrm{PX}^{\nu} = &\frac{1}{k_\nu}P_{\nu}  \mathcal{D}_\nu\theta_{\nu}\\ 
    &+P_{\nu} \Bigg[ \frac{1+\alpha_{x,\nu}^2}{2k_\nu^2w_{x,\nu}^2} +\frac{w_{x,\nu}^2}{4k_\nu}\mathcal{D}_\nu\left(\frac{\alpha_{x,\nu}}{w_{x,\nu}^2}\right)\\
    &\quad\quad\quad+\frac{k_{x,\nu}^2}{2 k_\nu^2}  -\frac{k_{x,\nu}}{k_\nu }\mathcal{D}_\nu x_{\mathrm{c},\nu} + x\rightarrow{y} \Bigg].
\end{split}
\end{equation}
where $\nu=1,\, 2,\, 3$ and $x\rightarrow y$ indicates the addition of identical terms with $x$ replaced by $y$. Because the pulses share a common propagation axis, the transverse integrals of ${\mathcal{L}}_\mathrm{NL}$ can be evaluated analytically, providing the reduced interaction Lagrangian:
\begin{equation}
\overline{\mathcal{L}}_{\mathrm{NL}} = -\frac{112\xi}{c}\; \left[P_{1}P_{2}\; \Gamma(1,2) + P_{1}P_{3}\; \Gamma(1,3)\right],
\end{equation}
where
\begin{equation}\label{eq:gamma}
\Gamma(\nu,\zeta) \equiv \frac{\exp{\left[ \frac{-2(x_{\mathrm{c},\nu}-x_{\mathrm{c},\zeta})^2}{w_{x,\nu}^2+w_{x,\zeta}^2} + \frac{-2(y_{\mathrm{c},\nu}-y_{\mathrm{c},\zeta})^2}{w_{y,\nu}^2+w_{y,\zeta}^2} \right]}}{\sqrt{(w_{x,\nu}^2+w_{x,\zeta}^2)(w_{y,\nu}^2+w_{y,\zeta}^2)}},
\end{equation} 
and $\nu$ and $\zeta$ label spectral components.
\begin{figure}[t]
\centering
\includegraphics[width=0.48\textwidth]{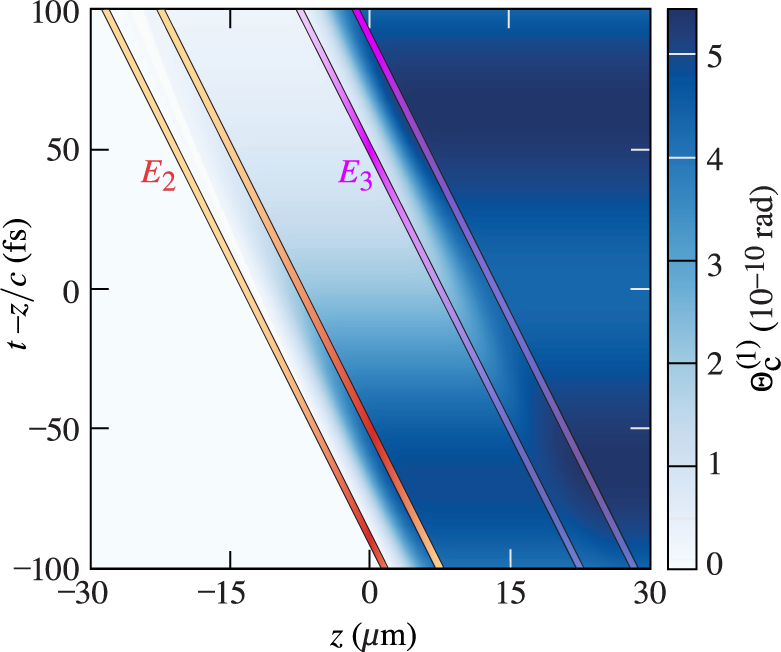}
\caption{Deflection angle $\Theta_\mathrm{c}^{(1)}$ of a probe pulse $E_1$, resulting from its  collision with two pump pulses, $E_2$ and $E_3$, as a function of $z$ and the probe moving-frame coordinate, $t-z/c$. All pulses have the same frequency $\hbar \omega =1.24$ eV, same minimum spot size $\tilde{w}_1=\tilde{w}_2 =\tilde{w}_3= 2$ $\mu$m, and share a focal plane at $z = 0$. Each pump has a peak power of $25$ PW and a duration of $20$ fs. The pump pulses are transversely offset by $x_{\mathrm{c},2}=x_{\mathrm{c},3}=1.4$ $\mu$m and are delayed by $\mp 70$ fs for $E_2$ and $E_3$, respectively, relative to the arrival time of the central time slice of the probe pulse at focus ($t-z/c = 0$, $z=0$). The temporal extent of the pump pulses are shown by the pairs of red and violet lines. The line brightness decreases with increasing distance from the focal plane, reflecting the intensity of the pumps.} \label{fig:3}
\end{figure}

With the reduced Lagrangian established, the principle of least action can be applied to derive equations of motion for the power, centroid, and spot size of each pulse. Variation with respect to $\theta_\nu$ reveals that the power of each pulse is conserved: $ \mathcal{D}_\nu P_\nu = 0$. Variation with respect to $\bvec{r}_{c,\nu}$ yields an equation describing the evolution of $\bvec{k}_{\perp,\nu}$ in response to nonlinear refraction. Recasting this result in terms of the centroid positions gives the following second-order equations:
\begin{equation}\label{eq:beamcent1}
\begin{split}
\mathcal{D}_1^2 x_{\mathrm{c},1} = 
- \frac{448\xi}{c}\; \Bigg[&\frac{(x_{\mathrm{c},1} - x_{\mathrm{c},2})}{(w_{x,1}^2 + w_{x,2}^2)^{1/2}}\;P_2\Gamma(1,2)\\
&+\frac{(x_{\mathrm{c},1} - x_{\mathrm{c},3})}{(w_{x,1}^2 + w_{x,3}^2)^{1/2}}\;P_3\Gamma(1,3)\Bigg],
\end{split}
\end{equation}
\begin{equation}\label{eq:beamcent2}
\mathcal{D}_2^2 x_{\mathrm{c},2} = 
- \frac{448\xi}{c}\; \frac{(x_{\mathrm{c},2} - x_{\mathrm{c},1})}{(w_{x,1}^2 + w_{x,2}^2)^{1/2}}\;P_1\Gamma(1,2),
\end{equation}
\begin{equation}\label{eq:beamcent3}
\mathcal{D}_3^2 x_{\mathrm{c},3} = 
- \frac{448\xi}{c}\; \frac{(x_{\mathrm{c},3} - x_{\mathrm{c},1})}{(w_{x,1}^2 + w_{x,3}^2)^{1/2}}\;P_1\Gamma(1,3), 
\end{equation}
with the corresponding equations for $y_{\mathrm{c},\nu}$ obtained by replacing $x$ with $y$. Equations~\eqref{eq:beamcent1}--\eqref{eq:beamcent3} show that each pulse is attracted to the highest intensity of the counter-propagating pulse(s). The factors involving the spot sizes account for the decrease in intensity due to diffraction, which weakens this attraction.  

Similarly, variations with respect to $w_{x,\nu}$ and $w_{y,\nu}$ yield equations that describe the evolution of the phase-front curvature parameters $\alpha_{x,\nu}$ and $\alpha_{y,\nu}$ in response to diffraction. Recasting these gives second-order equations for the spot sizes: 
\begin{widetext}
\begin{equation}\label{eq:spot1}
\mathcal{D}_1^2 w_{x,1} =\frac{4}{k_1^2 w_{x,1}^3} 
\Bigg\{ 1 - \frac{112\xi}{c} 
\frac{[(w_{x,1}^2 + w_{x,2}^2) - 4 (x_{\mathrm{c},1} - x_{\mathrm{c},2})^2]}{(w_{x,1}^2+w_{x,2}^2)^2 } \;
k_1^2 w_{x,1}^4 \,P_2\, \Gamma(1,2)+ 2 \rightarrow 3 
\Bigg\},
\end{equation}
where $2\rightarrow 3$ denotes the addition of an identical term with the labels  $2$ and $3$ exchanged, and
\begin{equation}\label{eq:spot2}
\mathcal{D}_2^2 w_{x,2} = \frac{4}{k_2^2 w_{x,2}^3} \Bigg\{ 1 - \frac{112\xi}{c} 
\frac{[(w_{x,1}^2 + w_{x,2}^2) - 4 (x_{\mathrm{c},1} - x_{\mathrm{c},2})^2]}{(w_{x,1}^2+w_{x,2}^2)^2 } \;
k_2^2 w_{x,2}^4 \, P_1\,\Gamma(1,2)
\Bigg\},
\end{equation}
\begin{equation}\label{eq:spot3}
\mathcal{D}_3^2 w_{x,3} = \frac{4}{k_3^2 w_{x,3}^3} 
\Bigg\{ 1 - \frac{112\xi}{c} 
\frac{[(w_{x,1}^2 + w_{x,3}^2) - 4 (x_{\mathrm{c},1} - x_{\mathrm{c},3})^2]}{(w_{x,1}^2+w_{x,3}^2)^2 } \;k_3^2 w_{x,3}^4 \,P_1\, \Gamma(1,3)
\Bigg\}.
\end{equation}
\end{widetext}
The corresponding equations for $w_{y,\nu}$ are obtained by replacing  $x$ with $y$. The first term on the right-hand side of Eqs. \eqref{eq:spot1}--\eqref{eq:spot3} (i.e., the 1 inside the brackets) gives rise to linear diffraction, while the subsequent terms contribute nonlinear diffraction or focusing due to the interaction between pulses.

At this point, the coupled system described by Eqs.~\eqref{eq:beamcent1}--\eqref{eq:spot3} can be integrated numerically to self-consistently determine the evolution of all pulses. However, the system simplifies considerably by treating the backward-propagating pulses, $E_2$ and $E_3$, as strong pumps, and the forward-propagating pulse, $E_1$, as a weak probe. In this approximation, all terms proportional to $P_1$ in the evolution equations for the pump parameters are neglected, and the pumps undergo linear diffraction with fixed centroids: 
\begin{equation}
    \mathcal{D}^2_2 x_{\mathrm{c},2}=0,\;\;
    \mathcal{D}^2_2 w_{x,2}=\frac{4}{k_2^2 w_{x,2}^3}, 
\end{equation}
and similarly for $x \rightarrow y$ or the subscript $2\rightarrow3$. By choosing $\bvec{k}_{\perp,2}=\bvec{k}_{\perp,3}= 0$ and setting the minimum spot size for both pumps in $x$ and $y$ to $\tilde{w}_2$ at $z= 0$, one recovers the familiar Gaussian optics result for their spot size evolution:
\begin{equation}\label{eq:LineDiffpump}  \bvec{r}_{\mathrm{c},2}=\bvec{r}_{0,2},\;\;w_{2} = \tilde{w}_2 \sqrt{1+\frac{4z^2}{k_2^2 \tilde{w}_2^4}},
\end{equation}
and similarly for the subscript $2\rightarrow3$, where $\bvec{r}_0$ is the fixed transverse location of the pump centroids.

In all scenarios relevant to laboratory experiments, the pump fields are significantly smaller than the Schwinger field, $E_\mathrm{S}$. This allows the coupled system of Eqs.~\eqref{eq:beamcent1} and \eqref{eq:spot1} to be solved perturbatively, with the light-pulse parameters of the probe expanded as follows: 
\begin{equation}
\begin{split}
    x_{\mathrm{c},1} &= x_{\mathrm{c},1}^{(0)} + x_{\mathrm{c},1}^{(1)} + \ldots, \\
    w_{x,1} &= w_{x,1}^{(0)} +  w_{x,1}^{(1)} + \ldots, 
\end{split}
\end{equation}
and similarly for $x \rightarrow y$. The zeroth-order terms, denoted by $(0)$, correspond to the initial centroid position and spot size resulting from linear diffraction, respectively. The first order terms, denoted by $(1)$, account for corrections of order ${\sim}\mathcal{O}(|E_2|^2/E_\mathrm{S}^2)$ and ${\sim}\mathcal{O}(|E_3|^2/E_\mathrm{S}^2)$.

To illustrate refraction of a probe pulse in the presence of the pumps, consider a probe with an unperturbed focal plane at $z=0$ and 
\begin{equation}\label{eq:LineDiffprobe}
    \bvec{r}_{\mathrm{c},1}^{(0)}=0, \;\;w_{x,1}^{(0)} = w_{y,1}^{(0)} = \tilde{w}_1 \sqrt{1+\frac{4z^2}{k_1^2 \tilde{w}_1^4}},
\end{equation}
where $\tilde{w}_1$ is the minimum  spot size. For pumps with $y_{\mathrm{c},1} = y_{\mathrm{c},2} = 0$ and $ w_2 = w_3$, the first-order correction to the probe centroid evolves according to
\begin{equation}\label{eq:beamcent1_pert}
\begin{split}
\mathcal{D}_1^2 x_{\mathrm{c},1}^{(1)} &= 
 \frac{448\xi}{c} \frac{1}{\sqrt{\big(w_{x,1}^{(0)}\big)^2+w_{2}^2}} \\ 
&\times\Big[x_{\mathrm{c},2}P_2\Gamma^{(0)}(1,2)
+ x_{\mathrm{c},3}P_3\Gamma^{(0)}(1,3)\Big],
\end{split}
\end{equation}
where $\Gamma^{(0)}$ denotes $\Gamma$ [Eq. \eqref{eq:gamma}] evaluated with the zeroth-order probe parameters. 

Figure \ref{fig:3} shows the solution of Eq. \eqref{eq:beamcent1_pert} for parameters motivated by planned high-power laser facilities. The solution is expressed in terms of the deflection angle 
\begin{equation}
\Theta_\mathrm{c}^{(1)} \equiv \mathrm{arctan}\left( \mathcal{D}_1 x_{\mathrm{c},1}^{(1)}\right)
\end{equation}
as a function of $z$ and the moving-frame coordinate of the probe $t-z/c$. The deflection angle is shown because, after the interaction, it is independent of propagation distance and therefore serves as a more general observable. In this example, all pulses have the same frequency, $\hbar\omega = 1.24$ eV, and minimum spot size, $\tilde{w}_1=\tilde{w}_2=\tilde{w}_3 = 2$~$\mu$m. Each pump pulse has a peak power of $25$ PW and a duration of $20$ fs. The pump pulses are transversely offset by $x_{\mathrm{c},2}=x_{\mathrm{c},3}=1.4$ $\mu$m and are delayed by $\mp70$~fs for $E_2$ and $E_3$, respectively, relative to the arrival of the probe's temporal center at focus ($t-z/c = 0$, $z=0$). The separation between the red and violet line pairs indicates the temporal extent of the pumps, while the line brightness represents their relative intensity.

The offset and timing of the pump pulses cause the centroid of the probe to be deflected upward, first toward the center of the earlier pump (red) and then toward the center of the later pump (violet). This is the same scenario illustrated in Fig. \ref{fig:1}(b). The central time slice of the probe, $t-z/c = 0$, encounters an equal intensity in both pumps: the first pump after it has passed through its focus, and the second pump an equal distance before its focus. Because both pumps are away from their focus, their peak intensities are lower, leading to a weaker net deflection. For earlier or later time slices, one pump is closer to its focus, resulting in a comparatively higher intensity and a stronger net deflection. 

\subsection{Birefringence}
The Euler--Heisenberg Lagrangian also predicts that vacuum is birefringent \cite{Martin, FFvacBire, ahmadiniaz2025towards,di2006light, heinzl2006observation, ReviewAntonino, ReviewKing, king2016vacuum, matheron2025simulating}. Similar to cross-phase modulation, this birefringence arises from Kerr-like nonlinearities $\mathcal{L}_\mathrm{NL}\propto-|E_{\beta j}|^2|E_{\gamma m}|^2$. The proportionality constant depends on the polarizations of the envelopes that compose the quartet (i.e., $j$ and $m$). As a result, orthogonal projections of a spectral component can acquire a relative phase difference during the interaction. For instance, an initially linearly polarized pulse could become elliptically polarized during its interaction with another pulse, as shown in Fig. \ref{fig:1}(c).    

To demonstrate the application of the reduced action integral approach to vacuum birefringence, consider two arbitrarily polarized pulses, $\beta = 1$ and $\gamma = 2$, propagating on the $z$-axis with $\hat{\mathbf{k}}_1 = -\hat{\mathbf{k}}_2 = \hat{\mathbf{z}}$. The action integral $\mathcal{S}$ for this configuration describes the evolution of eight coupled field envelopes: two co-polarized field envelopes, $(E_{11}, E_{21})$, their orthogonal counterparts, $(E_{12}, E_{22})$, and all respective complex conjugates. Among these, there are a total of eight phase-matched quartets. The interaction Lagrangian, 
\begin{equation}\label{eq:nonRedLBire}
\begin{split}
    \mathcal{L}_\mathrm{NL} = -&\xi\Bigg[\Big(14|E_{11}|^2|E_{22}|^2 +8|E_{11}|^2|E_{21}|^2\\
    +&14|E_{12}|^2|E_{21}|^2 +8|E_{12}|^2|E_{22}|^2\Big)\\
    -&3\Big( E_{11}^*E_{12}E_{21}^* E_{22} +E_{11}E_{12}E_{21}^*E_{22}+\mathrm{c.c.}\Big)\Bigg],
    \end{split}
\end{equation}
features two types of couplings that  determine the polarization dynamics: Kerr-like terms and ``polarization-scattering'' terms, corresponding to the first and second contributions grouped by parentheses, respectively. The Kerr-like terms exhibit two distinct coupling strengths that depend on the polarization orientation of the coupled envelopes. These terms are responsible for vacuum birefringence and result in orthogonally polarized components acquiring a relative phase difference. The polarization-scattering terms describe processes in which photons in one polarization state are scattered into the orthogonal state, e.g., $E_{11}\rightarrow E_{12}^*$. These terms cause changes in the polarization angle. 

The effect of birefringence (i.e., the Kerr-like terms) can be isolated by considering a pump--probe geometry, where a strong, linearly polarized pump ($\gamma = 2$) collides head-on with a weak, arbitrarily polarized probe ($\beta = 1$). In this geometry, the envelope $E_{22} = 0$ while all other envelopes are non-zero. This simplifies the interaction Lagrangian to
\begin{equation}\label{eq:simpBire}
    \mathcal{L}_\mathrm{NL} = -\xi \left(C_1|E_{11}|^2|E_{21}|^2 +C_2 |E_{12}|^2|E_{21}|^2\right), 
\end{equation}
where the coefficients $C_1=8$ and $C_2=14$ reveal the relative strength of the coupling to each polarization projection of the probe.

If the two pulses collide in a symmetric configuration with no lateral offsets, the lowest-order LG mode can be used for their trial functions:
\begin{equation}\label{eq:lowestLG}
\begin{split}
    &E = A\, \mathrm{exp}\left[-(1-i\alpha)\frac{r^2}{w^2}+i\ell \phi+i\theta\right].
\end{split}
\end{equation}
The components of the reduced Lagrangian are then 
\begin{equation}
\begin{split}
\overline{\mathcal{L}}_\mathrm{PX}^{\nu o} &= \frac{1}{k_\nu}P_{\nu o}  \mathcal{D}_\nu\theta_{\nu o}\\
    &+P_{\nu o} \left[ \frac{1+\alpha_{\nu o}^2}{k_\nu^2w_{\nu o}^2} +\frac{w_{\nu o}^2}{2k_\nu}\mathcal{D}_\nu\left(\frac{\alpha_{\nu o}}{w_{\nu o}^2}\right) \right],
    \end{split} 
\end{equation}
\begin{equation}
\overline{\mathcal{L}}_\mathrm{NL}= -\frac{8C_1\xi}{c} \frac{ P_{11}P_{21}}{w_{11}^2+w_{21}^2} -\frac{8C_2\xi}{c} \frac{ P_{12}P_{21}}{w_{12}^2+w_{21}^2}, 
\end{equation}
where $(\nu,o)$=(1,1), (1,2), or (2,1).
Variations of $\overline{\mathcal{L}}_\mathrm{NL}$ with respect to the light-pulse parameters of the pump yield terms that are proportional to the probe powers. Because the probe is weak, these terms can be approximated as zero. The light-pulse parameters of the pump are then the usual expressions for a linearly propagating Gaussian beam. Here, the pump is focused to a minimum spot size $\tilde{w}_2$ in the plane $z=0$, such that its spot size evolves according to $w_{21} = \tilde{w}_2 [1+4z^2/(k_2^2 \tilde{w}_2^4)]^{1/2}$.

\begin{figure}[t]
\centering
\includegraphics[width=0.48\textwidth]{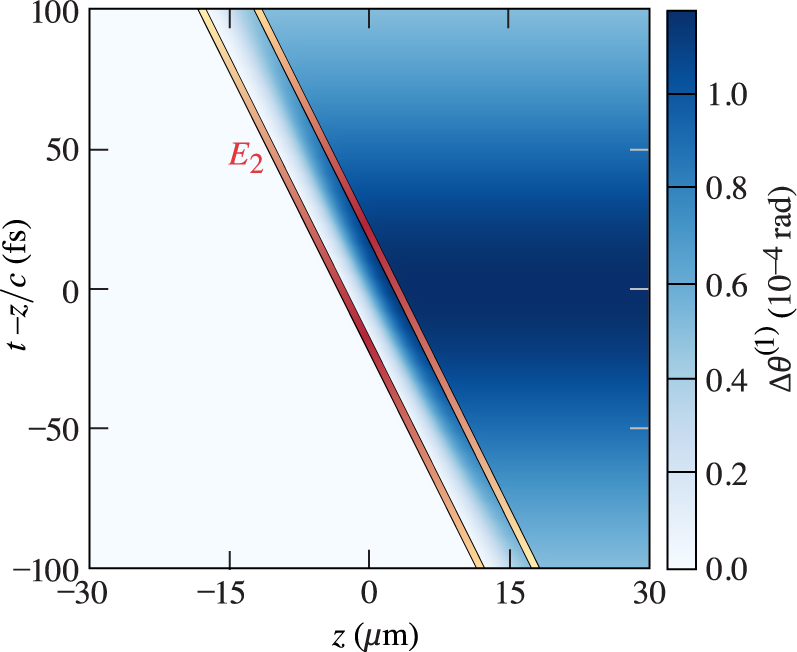}
\caption{The relative phase difference, $\Delta \theta^{(1)}_1$, between two polarization components of an x-ray probe pulse, $E_{11}$ and $E_{12},$ due to its interaction with a counter-propagating optical pump pulse, $E_2$, as a function of $z$ and the moving-frame coordinate of the probe $t-z/c$. The x-ray probe has a central frequency $\hbar\omega_1 = 25$ keV and a minimum spot size $\tilde{w}_1  = 1$ $\mu$m. The optical pump has a central frequency $ \hbar\omega_2 = 1.24$ eV, a minimum spot size $\tilde{w}_2 = 2$ $\mu$m, a peak power of $25$ PW, and a duration of $20$ fs. The pulses share a focal point in the plane $z=0$. The two red lines mark the front and back edges of the pump. The decreasing line brightness with increasing distance from the focal plane conveys the local intensity of the pump.} \label{fig:4}
\end{figure}

Varying the total reduced Lagrangian with respect to $P_{1j}$ provides equations of motion for the $\theta_{1j}$:
\begin{equation}\label{eq:phaseEvolution}
\mathcal{D}_1 \theta_{1j} = -\frac{2}{k_1 w_{1j}^2}+ \frac{8 C_j \xi k_1 P_2}{c} \frac{2w_{1j}^2+w_{21}^2}{(w_{1j}^2+w_{21}^2)^2}.
\end{equation}
This result is consistent with the expression derived using the moment method (Eq. C9 of Ref. \cite{Martin}).

Birefringence is characterized by the accumulated phase difference between the two polarization envelopes of the probe, $\Delta \theta_1 = \theta_{12} - \theta_{11}$. As in the previous example, the smallness of the nonlinearity justifies a perturbative approach to solving for $\Delta \theta_1$. That is, when integrating $\mathcal{D}_1 \Delta\theta_1$, the effect of the pump on the spot size of the probe can be neglected to leading order. Setting the focal plane of the probe at $z=0$ and using identical spot sizes for each polarization component, $w_{11}^{(0)} = w_{12}^{(0)} = \tilde{w}_1 [1+4z^2/(k_1^2 \tilde{w}_1^4)]^{1/2}$, provides a simplified equation for the evolution of the first-order phase difference:
\begin{equation}\label{eq:bire}
\mathcal{D}_1 \Delta\theta^{(1)}_1 = \frac{48 \xi k_1 P_2}{c} \frac{2\big(w_{11}^{(0)}\big)^2+w_{21}^2}{\big[\big(w_{11}^{(0)}\big)^2 +w_{21}^2\big]^2}.
\end{equation}

Figure \ref{fig:4} shows the solution to Eq. \eqref{eq:bire} for the collision of an x-ray probe pulse with a high-power optical pump pulse. The solution is plotted as a function of $z$ and the moving-frame coordinate of the probe $t-z/c$, with the temporal center of the probe occurring at $t-z/c = 0$. The probe pulse has a minimum spot size $\tilde{w}_1 = 1$ $\mu$m and a central frequency $\hbar\omega_1 = 25$ keV, selected to enhance the birefringence [the right-hand-side of Eq. \eqref{eq:bire} is proportional to $k_1$]. The pump, whose leading and trailing edges are demarcated by the red lines, has a peak power of $25$ PW, a pulse duration of $20$ fs, a minimum spot size $\tilde{w}_2 = 2$ $\mu$m, and a frequency $\hbar\omega_2 =1.24$ eV. The shading of the red lines indicates the intensity of the pump, which drops away from the focal plane ($z=0$) due to diffraction. Earlier or later time slices of the probe interact with the pump down or upstream from the focal plane, where its intensity is lower. This reduces the phase difference accumulated in these time slices.

\subsection{Frequency Mixing}
Electromagnetic waves propagating through a nonlinear medium undergo frequency mixing, generating distinct spectral components. The Euler--Heisenberg Lagrangian extends this hallmark process of classical nonlinear optics to vacuum. A number of theoretical and numerical studies have explored possible configurations for frequency mixing in vacuum \cite{Lundin, PhysRevA.98.023817, lundstrom2006using,PhysRevA.98.023817, ReviewAntonino, grismayer2021quantum, ReviewKing, ZhangAddsSemiClassical}, with the configuration proposed by Lundstr{\"o}m \textit{et al.} \cite{lundstrom2006using,Lundin,Hans&Antonino} providing the basis for a flagship experiment on the planned NSF OPAL laser \cite{nsf_opal_flagships}. The reduced=action-integral approach complements these efforts by providing equations for measurable quantities, such as the spot size, power, and photon number of the generated component.

Recall that $\mathrm{X}^{(3)}$ vanishes for co-propagating fields, which limits the interaction geometries capable of generating a new, spectrally distinct signal. In order to generate such a signal, the configuration proposed by Lundstr{\"o}m \textit{et al.} collides three pump pulses with mutually orthogonal propagation axes and polarizations, as depicted in Fig. \ref{fig:1}(d) \cite{lundstrom2006using, Lundin}. The relevant quartets include a fundamental pump envelope $E_{32}^*$ with $\unitvec{k}_3=\unitvec{z}$; two second-harmonic pump envelopes $E_{13}$ and $E_{21}$ with  $\unitvec{k}_1=\unitvec{x}$ and $\unitvec{k}_2=\unitvec{y}$; and the generated third-harmonic $E_{4o}^*$ with $\unitvec{k}_4 = \tfrac{1}{3}(2\unitvec{x}+2\unitvec{y}-\unitvec{z})$. For notational brevity, the polarization labels of the pumps will be omitted: $E_{13}\rightarrow E_1$, $E_{21}\rightarrow E_2$, and $E_{32}^*\rightarrow E_3^*$. The interaction Lagrangian for the signal generation is then given by
\begin{equation}
    \mathcal{L}_\mathrm{NL} = -\frac{7\xi}{2} E_1E_2E_3^*\left(\tfrac{2}{3}E_{41}^* -\tfrac{1}{3}E_{42}^* +\tfrac{2}{3} E_{43}^*  \right) +\mathrm{c.c.},
\end{equation}
where $\bvec{E}_4$ has three projections in the coordinate system aligned to the pump propagation axes, i.e., $E_{4o}$ for $o = 1,2,3$. 

Generation of the third-harmonic signal only occurs within the overlap volume of the pumps. The largest signal will be produced when all three pumps share a focal point, such that the overlap volume is determined by their spot sizes. Within this volume, the spot sizes of the pumps can be approximated as constant and their envelopes modeled as
\begin{equation}\label{eq:inputform}
    E_\beta =  \sqrt{\frac{16\tilde{P}_\beta}{c\tilde{w}_\beta^2}}\,\mathrm{exp}\left[ -\frac{(z_\beta - c t)^2}{c^2\tau_\beta^2} -\frac{r_\beta^2}{\tilde{w}_\beta^2}\right]
\end{equation}
for $\beta = 1,\,2,\,3$, where $\tilde{P}_\beta$ is the peak power, $\tilde{w}_\beta$ is the minimum spot size, and $\tau_\beta$ is the pulse duration. This expression for the pump envelopes neglects the effects of the pumps on each other. For existing or planned laser parameters, these effects would result in minute corrections to $\mathcal{L}_\mathrm{NL}$ of order ${\sim}\mathcal{O}(|E_{\beta}|^2/E_\mathrm{S}^2)$.



When the pumps have different minimum spot sizes, their overlap volume is ellipsoidal. To simplify calculation of the reduced Lagrangian in this volume, a rotated coordinate system $(x_4,\,y_4,\,z_4)$ is adopted, where $z_4$ is the propagation axis of the signal and $(x_4,\,y_4)$ are aligned to the major and minor axes of the pump-overlap ellipse in the $z_4 = 0$ plane. This reduces $\bvec{E}_4$ to two projections---$o' = 1$ parallel to $x_4$ and $o' = 2$ parallel to $y_4$. The evolution of these projections in the asymmetric volume of the pumps is modeled using the lowest-order HG mode for their trial functions:
\begin{equation}\label{eq:E4HGTrial}
\begin{split}
    &E_{4o'} = A\,\mathrm{exp}\big[i\theta +i\bvec{k}_\perp\cdot\big(\bvec{r}-\bvec{r}_\mathrm{c}\big)\big]\\
                  & \times \exp{\left[-(1-i \alpha_x)\frac{(x-x_\mathrm{c})^2}{w_x^2}-(1-i \alpha_y)\frac{(y-y_\mathrm{c})^2}{w_y^2}\right]},
\end{split}
\end{equation}
where the label $4o'$ on each light-pulse parameter is suppressed for conciseness. 

Before applying the reduced action integral to the ellipsoidal geometry, it is instructive to first consider a simpler geometry where each pump has the same minimum spot size, $\tilde{w}_1=\tilde{w}_2=\tilde{w}_3= \tilde{w}_0$. In this case, the overlap volume is spherical, allowing the polarization of the signal to be aligned with one of the transverse axes, $x_4$ or $y_4$. The signal can then be represented by a single trial function for the envelope
\begin{equation}
E_{41'} = \tfrac{2}{3}E_{41} -\tfrac{1}{3}E_{42} +\tfrac{2}{3}E_{43},
\end{equation}
which will be renamed $E_4$ for brevity. 

Owing to the spherical symmetry of the overlap volume, the signal pulse will be cylindrically symmetric about its propagation axis and have a fixed centroid. As a result, $w_{x,4} = w_{y,4} \equiv w_4$ and $\bvec{r}_{\mathrm{c},4} = 0$, so that the trial function is the lowest-order LG mode. The reduced paraxial Lagrangian is then
\begin{equation}
\overline{\mathcal{L}}_\mathrm{PX}^{4} = \frac{1}{k_4}P_{4}  \mathcal{D}_4\theta_{4} +P_{4} \left[ \frac{1+\alpha_{4}^2}{k_4^2w_{4}^2} +\frac{w_{4}^2}{2k_4}\mathcal{D}_4\left(\frac{\alpha_{4}}{w_{4}^2}\right) \right].
\end{equation}
The reduced interaction Lagrangian can be simplified by considering an overlap volume with an extent that is smaller than the length of the pumps: $\tilde{w}_0 <c\tau_\beta$. In this regime, the $z_\beta$ dependence of Eq. \eqref{eq:inputform} can be neglected, and the pump powers treated as functions of time alone: $P_\beta \approx \tilde{P}_\beta \,\mathrm{exp}(-2t^2/\tau_\beta^2)$. The resulting reduced interaction Lagrangian is 
\begin{equation}
\begin{split}
 \overline{\mathcal{L}}_\mathrm{NL} 
   = -\frac{56 \xi }{ c  }\frac{w_4(P_1P_2P_3P_4)^{1/2}\exp{\left(-\frac{2z_4^2}{\tilde{w}_0^2}-i\theta_4\right)}}{\tilde{w}_0^3(1+i\alpha_4)+2 \tilde{w}_0 w_4^2} +\mathrm{c.c.}
\end{split}
\end{equation}
Unlike the previous examples, $\overline{\mathcal{L}}_\mathrm{NL}$ depends explicitly on a phase, appearing here through the signal phase $\theta_4$.
This dependence indicates that the power of the signal ($P_4$) is not conserved, consistent with its generation by four-wave mixing. 

Varying the total reduced Lagrangian with respect to $\theta_4$ provides the equation of motion for the signal power
\begin{equation}\label{eq:gendPowdz}
\begin{split}
    &\mathcal{D}_4 P_4 =  \frac{56 \xi k_4}{ c  }\\
    &\times\frac{w_4(P_1P_2P_3P_4)^{1/2}\exp{\left[-\frac{2z_4^2}{\tilde{w}_0^2}-i(\theta_4 -\frac{\pi}{2})\right]}}{\tilde{w}_0^3(1+i\alpha_4)+2 \tilde{w}_0 w_4^2} +\mathrm{c.c.}
\end{split}
\end{equation}
Equation \eqref{eq:gendPowdz} along with the equations of motion for $\alpha_4$, $w_4$, and $\theta_4$ constitute a system of four coupled first-order differential equations. These equations can be solved numerically or analytically by assuming that diffraction is negligible over the signal formation length, i.e., $\mathcal{D}_4 w_4 \approx 0$ ($\alpha_4 \approx 0$) and $\mathcal{D}_4 \theta_4 \approx 0$. This assumption is typically justified because the signal pulse is generated over the length of the overlap volume ${\sim} \tilde{w}_0$, whereas appreciable diffraction occurs over a longer distance ${\sim} k_4\tilde{w}_0^2$. Under these conditions, the spot size and phase of the signal within the overlap volume are $w_4= \tilde{w}_0/\sqrt{2}$ and $\theta_4 =\pi/2$. Using these expressions, Eq. \eqref{eq:gendPowdz} can be integrated for pumps with equal durations $\tau_1=\tau_2=\tau_3 = \tau$ and the boundary condition $P_4\rightarrow 0$ as $z_4\rightarrow -\infty$ to yield
\begin{equation}\label{eq:P4}
\begin{split}
    P_4 &= \frac{49 \pi \xi^2  k_4^2T^2\,\tilde{P}_1 \tilde{P}_2 \tilde{P}_3}{ \tilde{w}_0^2(c^2T^2+\tilde{w}_0^2)} \;\exp{\left[\frac{-4(z_4 - c t)^2}{c^2T^2+\tilde{w}_0^2}\right]}\\
    & \times\left[ 1+ \mathrm{erf}\left(\frac{c^2 T^2z_4+\tilde{w}_0^2ct}{\tfrac{1}{\sqrt{2}}c\tilde{w}_0T \sqrt{c^2T^2+\tilde{w}_0^2}}\right)\right]^2,
    \end{split}
\end{equation}
where  $\mathrm{erf}(z)$ is the error function and $T^2 \equiv 2 \tau^2/3$. In the far field ($z_4\gg\tilde{w}_0$), the duration and spectral bandwidth of the signal pulse are characterized by $(T^2 + \tilde{w}_0^2/c^2)^{1/2}$ and its reciprocal, respectively.

The total number of signal photons $N_4$ can be estimated by integrating the far-field limit ($z_4 \rightarrow \infty$) of Eq. \eqref{eq:P4} over time and dividing by the photon energy $\hbar \omega_4$. This provides
\begin{equation}\label{eq:NumPhots}
    N_4 \approx \frac{98 \pi^{3/2}\xi^2   k_4\, \tilde{P}_{1} \tilde{P}_{2} \tilde{P}_{3} }{  \hbar c^2}\;\frac{T^2}{\tilde{w}_0^2\sqrt{c^2T^2+\tilde{w}_0^2}}.
\end{equation}
This result agrees with that obtained using a Green's function approach (Eq.~(8) of Ref.~\cite{lundstrom2006using}), up to differences in approximations for the overlap volume and temporal profiles of the pump pulses. 

\begin{figure*}[t]
\centering
\includegraphics[width=1\textwidth]{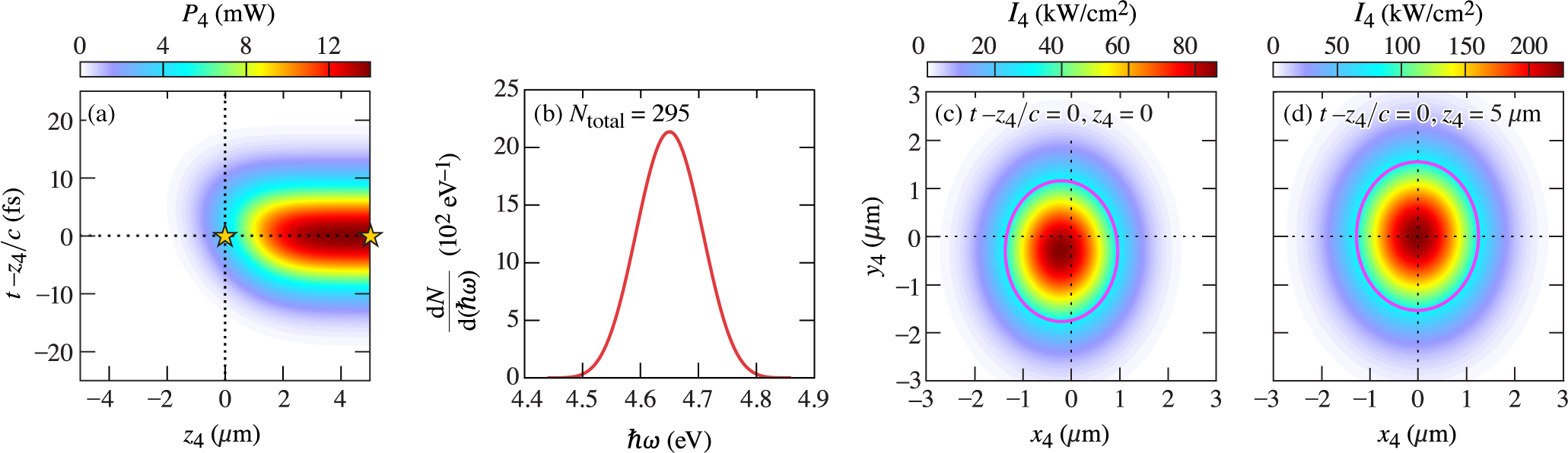}
\caption{Generation of a third-harmonic signal pulse, $E_4$ ($\hbar \omega_4 = 4.65$ eV), from the interaction of two second-harmonic pump pulses, $E_1$ and $E_2$ ($\hbar \omega_{1}=\hbar \omega_{2} = 3.1$ eV), and a fundamental pump pulse, $E_3$ ($\hbar \omega_3 = 1.55$ eV). The pumps have unequal spot sizes $\tilde{w}_1=2$ $\mu$m, $\tilde{w}_2=3$ $\mu$m, and $\tilde{w}_3=5$ $\mu$m, equal durations $\tau_1=\tau_2=\tau_3=20$ fs, and peak powers $4\tilde{P}_1=4\tilde{P}_2=\tilde{P}_3 = 25$ PW. (a) Power of the generated signal, $P_4$, along the optical axis, $z_4$, as a function of its moving-frame coordinate $t-z_4/c$. (b) Spectral density of the signal at a $z_4$ location beyond the overlap volume, where its generation has ceased; integration gives a total of $295$ third-harmonic photons. (c)  At the location of maximum power growth [center star in (a) at $t-z_4/c = 0$ fs and $z_4= 0$ $\mu$m], the different spot sizes of the pumps results in a signal with an elliptical transverse profile (outlined in violet) and an off-axis centroid. (d) After the signal generation ceases [right star in (a) at $t-z_4/c = 0$ fs and $z_4 = 5$ $\mu$m], the centroid moves to the origin.} \label{fig:5}
\end{figure*}

Figure \ref{fig:5} presents the results of the reduced action approach for a more complex scenario in which pumps with different spot sizes create an ellipsoidal overlap volume with $c\tau < \tilde{w}_\beta$. As in the phase modulation example (Sec. \ref{sec:Apps} A), the parameters are chosen to approximate the anticipated parameters of the planned NSF OPAL facility: $\tilde{w}_1=2$ $\mu$m, $\tilde{w}_2=3$ $\mu$m, $\tilde{w}_3=5$ $\mu$m, and $\tau=20$ fs \cite{nsf_opal_capabilities}.  NSF OPAL will natively generate two pulses at its fundamental frequency; therefore, one of the fundamental pulses must be frequency-doubled and subsequently split to achieve the desired interaction geometry [Fig. \ref{fig:1}(d)]. Accounting for frequency conversion losses and pulse splitting, the peak powers of the pumps are set to $4\tilde{P}_1=4\tilde{P}_2=\tilde{P}_3 = 25$ PW. 

The ellipsoidal overlap volume necessitates the use of Eq.~\eqref{eq:E4HGTrial} for the trial functions of the two signal-pulse projections ($o' = 1$ and $o' = 2$). In addition, due to the ultrashort pump durations, the $z_\beta$ dependence of the pump powers must be retained [see exponent of Eq. \eqref{eq:inputform}]. Thus, solving for the evolution of the signal-pulse parameters requires numerically integrating a system of 20 coupled first-order partial differential equations---one for each of the ten parameters in the two projections.  

Figure \ref{fig:5}(a) shows the spatiotemporal evolution of the signal power as a function of $z_4$ and its moving-frame coordinate $t-z_4/c$. The peak power of all three pumps simultaneously reach their focus at $t-z_4/c = 0$ and $x_4 = y_4 = z_4 = 0$, resulting in the strongest signal generation at $t-z_4/c = 0$. The photon spectrum displayed in Fig. \ref{fig:5}(b) was obtained from the Fourier transform of the time-domain electric field at a $z_4$ location beyond the overlap volume, where  signal generation has ceased. Integration over the spectrum yields a total of $295$ third-harmonic photons ($\hbar \omega_4=4.65$ eV). For comparison, Eq. \eqref{eq:NumPhots} provides an estimate of $376$ third-harmonic photons when evaluated with the mean spot size of the pumps ($\tilde{w}_0 = 3.33$ $\mu$m). 

The signal exhibits an elliptical transverse profile and an evolving centroid, as shown in Fig. \ref{fig:5}(c) at the location of maximum power growth and in Fig. \ref{fig:5}(d) at a location where the power growth has ceased. These two locations correspond to the center and right stars in Fig. \ref{fig:5}(a), respectively. The initial displacement of the centroid from the transverse origin ($x_4 = y_4 = 0$) arises from asymmetric generation of the signal within the ellipsoidal volume. For larger $z_4$, the signal is generated more strongly near the origin, causing the centroid to move toward the $z_4$ axis.

\section{Conclusions and Outlook}\label{sec:Con}
A reduced action integral was developed to model the nonlinear interaction of light pulses in vacuum. The coupling between pulses arises from their mutual polarization and magnetization of virtual electron--positron pairs, as predicted by the Euler--Heisenberg Lagrangian. Starting from the full inhomogeneous wave equations describing all nonlinear interactions, the slowly varying envelope and paraxial approximations are used to construct a simplified action integral for interactions among spectrally distinct field components. A trial function for each component, parameterized in terms of intuitive quantities like the spot size, power, and phase, is substituted into the action integral. This substitution allows integration over the coordinates transverse to the propagation direction of each field component, yielding a reduced action integral. The strength of this approach lies in the fact that the principle of least action provides equations of motion for experimentally observable quantities. Consequently, this method enables rapid exploration of potential geometries and parameters prior to performing more intensive simulations or experiments. The approach was applied to three examples, phase modulation, birefringence, and frequency mixing, demonstrating its versatility.

The examples presented in this work consider relatively simple geometries, in which the nonlinear coupling $\mathcal{L}_\mathrm{NL}$ can be reduced to an analytic expression, $\overline{\mathcal{L}}_\mathrm{NL}$, in terms of the light-pulse parameters. The requirements for this reduction are that the coupled pulses either share a common propagation axis or have trivial transverse dependence. More elaborate geometries can be broadly divided into two categories: (1) those involving strong pumps and weak probes, where the pumps affect the probes but the probes have no effect on other pulses, and (2) those in which all pulses affect each other.

In the first case, the reduced Lagrangian $\overline{\mathcal{L}}_\mathrm{NL}$ can be found by numerically integrating over the transverse coordinates after substituting discretized trial functions into $\mathcal{L}_\mathrm{NL}$. The advantage of numerical integration is that it allows for non-paraxial trial functions (as long as $\mathcal{L}_\mathrm{PX}$ is replaced by its non-paraxial counterpart $\mathcal{L}_\mathrm{NPX}$), which can improve accuracy, especially in geometries with tight focusing \cite{Limitations}. Since the transverse integration commutes with the variational derivatives, the only computational effort required lies in performing the integrals themselves. That is, $\mathcal{L} = \mathcal{L}_\mathrm{(N)PX} + \mathcal{L}_\mathrm{NL}$ only contains derivatives of analytic functions and is therefore analytic itself. 

In the second case, where all pulses affect each other, the transverse integrals cannot generally be taken, since doing so would require prior knowledge of the pulse evolution at future times. Nevertheless, because the vacuum nonlinearity is extremely small, a perturbative approach is almost always justified. To leading order, this reduces the second case to the first. Then, if required, higher-order corrections can be included iteratively. 

The computational simplicity of the reduced-action-integral approach positions it well for gradient-based optimization. For instance, the input parameters of pump or probe pulses can be optimized to maximize a desired observable effect of vacuum nonlinearity. This effect can be encoded in a loss function, whose derivatives with respect to the parameters determine how they are updated at each iteration of the optimization loop. Leveraging automatic differentiation, which is readily available in modern software libraries, could dramatically accelerate this process, particularly in geometries with many input parameters \cite{Miller2025Spatiotemporal}. This is a promising direction for future work, with the ultimate goal of detecting the effects of photon--photon scattering in the interaction of electromagnetic waves. 

\medskip
\begin{acknowledgments}
The authors thank A. Di Piazza, H. G. Rinderknecht, B. Barbosa, K. G. Miller, A. Konzel,  A. L. Elliott, and L. S. Mack for helpful and insightful discussions.

This report was prepared as an account of work sponsored by an agency of the U.S. Government. Neither the U.S. Government nor any agency thereof, nor any of their employees, makes any warranty, express or implied, or assumes any legal liability or responsibility for the accuracy, completeness, or usefulness of any information, apparatus, product, or process disclosed, or represents that its use would not infringe privately owned rights. Reference herein to any specific commercial product, process, or service by trade name, trademark, manufacturer, or otherwise does not necessarily constitute or imply its endorsement, recommendation, or favoring by the U.S. Government or any agency thereof. The views and opinions of authors expressed herein do not necessarily state or reflect those of the U.S. Government or any agency thereof.

This material is based upon work supported by the Department of Energy [National Nuclear Security Administration] University of Rochester “National Inertial Confinement Fusion Program'' under Award Number DE-NA0004144. The work of M. S. F. is supported by the European Union’s Horizon Europe research and innovation program under the Marie Sklodowska-Curie Grant Agreement No. 101105246STEFF.
\end{acknowledgments}

\appendix
\section{Expression for $\mathrm{X}^{(3)}$}
The third-order symmetrized susceptibility tensor governing nonlinear interactions in vacuum can be written as
\begin{equation}
\begin{split}
    &\mathrm{X}^{(3)}_{jmno}(\beta,\gamma,\mu,\nu) = \frac{\xi}{64\pi } \sum_{\{(\beta j),(\gamma m),(\mu n),(\nu o)\} }\\\bigg\{ &2[(1-\unitvec{k}_\gamma\cdot\unitvec{k}_\mu)\delta_{mn}+\hat{k}_{\gamma, n}\hat{k}_{\mu, m}][(1-\unitvec{k}_\nu\cdot\unitvec{k}_\beta)\delta_{oj} ] \\
    +&2[(1-\unitvec{k}_\gamma\cdot\unitvec{k}_\mu)\delta_{mn}+\hat{k}_{\gamma, n}\hat{k}_{\mu, m}](\hat{k}_{\nu, j}\hat{k}_{\beta, o} ) \\
    +&7[\unitvec{k}_\mu\cdot(\unitvec{k}_\nu-\unitvec{k}_\beta)](\delta_{mj}\delta_{on}-\delta_{nj}\delta_{om})\\
    +&7\hat{k}_{\mu, j}[(\hat{k}_{\nu, n}-\hat{k}_{\beta, n})\delta_{om}-(\hat{k}_{\nu, m}-\hat{k}_{\beta, m})\delta_{on}]\\
    +&7\hat{k}_{\mu, o}[(\hat{k}_{\nu, m}-\hat{k}_{\beta, m})\delta_{nj}-(\hat{k}_{\nu, n}-\hat{k}_{\beta, n})\delta_{mj}] \bigg\},
    \end{split}
\end{equation}
where the notation $\unitvec{k}_{\zeta, l}$ denotes the $l^{\mathrm{th}}$ Cartesian projection of $\unitvec{k}_\zeta$ and $\delta$ is the Kronecker delta. In the summation, the members of a phase-matched quartet, e.g., $(\beta j),(\gamma m),(\mu n),(\nu o)$, are fixed. Therefore, the summation is taken over all $4!$ permutations of the tuples $\{(\beta j),(\gamma m),(\mu n),(\nu o)\}$. That is, each permutation acts simultaneously on the component label $(\beta,\gamma,\mu,\nu)$ and polarization-projection label $(j,m,n,o)$, preserving the full envelope label.  For example, exchanging $(\beta j)$ and $(\gamma m)$ in a permutation means both $\beta \leftrightarrow \gamma$ and $j \leftrightarrow m$ simultaneously. Note that despite the apparently antisymmetric appearance of the terms with 7 as a coefficient, they do not vanish under symmetrization. The sign change introduced by interchanging the component labels of the wavevectors is compensated by the sign change from interchanging polarization labels. 

\clearpage
\bibliographystyle{apsrev4-1}
\bibliography{VAR}
\end{document}